\documentclass{elsarticle}
\usepackage{graphicx} % Required for inserting images
\usepackage{amsmath}
\usepackage{amssymb}
\usepackage{amsfonts}
\usepackage{natbib}
\usepackage{IEEEtrantools}
\usepackage{appendix}
\newcommand{\ai}{\emph{ab initio}}
\newcommand{\EinA}{Einstein \emph{A} coefficients}
\newcommand{\etal}{\emph{et al.}}
\usepackage{color}

\biboptions{sort&compress}

\title{The predictive power of the calculated line lists of carbon monoxide}
\author{E. S. Medvedev}
\author{V. G. Ushakov}
\address{Federal Research Center of Problems of Chemical Physics and Medicinal Chemistry (former Institute of Problems of Chemical Physics), Russian Academy of Sciences, 142432 Chernogolovka, Russian Federation}

\date{October 2025}

\begin{document}
\begin{abstract}

The ability of our semi-empirical irregular dipole-moment functions (2022) and (2025) to predict the intensities of the yet unobserved lines, as well as to describe the observed ones not used in the fitting, is demonstrated by comparison with recent measurements in the 0-0, 1-0, 3-0, and 7-0 bands.

\end{abstract}

\maketitle

% \section{Introduction}

The spectroscopic databases such as HITRAN \cite{Gordon22}, ExoMol \cite{Tennyson17}, and others collect the line parameters calculated with use of the molecular functions derived from the experimental and \ai\ data. The data cover not only the observed lines but also yet unobserved ones. Therefore, the first requirement is good description of the input data and the second one to make meaningful predictions. The second task is very challenging because the form of the molecular functions may strongly affect the calculated results \cite{Medvedev15,Medvedev16,Medvedev21,Medvedev22OC}, as was first demonstrated in the paper by Li \etal\ \cite{Li15} where the high-overtone lines were excluded from the list.

We have published three line lists for CO calculated with the semi-empirical potential-energy function of Meshkov \etal\ \cite{Meshkov18} and various semi-empirical dipole-moment functions (DMFs), namely regular (\emph{i.e.}, single-valued, having no branch points in the complex plane) in Ref. \cite{Meshkov22} and irregular (many-valued, with branch points) in Refs. \cite{Medvedev22,Ushakov25CO}; the rational DMF (the ratio of two polynomials) was also tried in Ref. \cite{Medvedev22}, which gave the intensities close to those obtained with the irregular DMF, therefore the differences can serve as a qualitative estimate of the precision of the calculation. 

In Ref. \cite{Meshkov22}, we collected all the experimental data available at that time and used them along with our own \ai\ calculated DMF to fit the regular DMF and to create line list 1 (LL1). The LL1 was restricted by transitions with $\Delta v\le6$ since higher overtone-transition intensities were considered to be unreliable. In Ref. \cite{Medvedev22}, we used the same input dataset to fit the irregular DMF and to create line list 2 (LL2), which was not subject to the above restriction in $\Delta v$. For comparison, we used the rational DMF and found moderate differences with LL2, which we treated as qualitative estimates of the calculation accuracy. In particular, we found for the 7-0 band the error estimate to be about 30\%. It is of interest to verify these predictions with new data.

Later, the 7-0 band has been measured by Balashov \etal\ \cite{Balashov23} and new high-precision 3-0 band intensities have been published by Bielska \etal\ \cite{Bielska22} and Hodges \etal\ \cite{Hodges25}. We tried to describe these data using our irregular DMF and found that they came in contradiction with the earlier data on the 1-0 band \cite{Zou02,Devi18} since the 1-0 data contained systematic errors. Therefore, the 1-0 data were excluded from the input dataset, and the above data on the 3-0 and 7-0 bands were included to calculate line list 3 (LL3). In particular, the LL3 contains predictions for the 1-0 band, which can be verified now.

Recently, more measurements have been performed with extremely high precision by Bailey \etal\ \cite{Bailey25CO} in the 1-0 band, Huang \etal\ \cite{Huang24}, Li \etal\ \cite{Li25ratios30}, and Cygan \etal\ \cite{Cygan25} in the 3-0 band, and high-precision measurements in the 0-0 band by Tretyakov \etal\ \cite{Tretyakov23} were also available; all these data, except for the 7-0 band, were not used in calculations of LL3. 

The purpose of this publication is to compare the newly observed line intensities, as well as the earlier data not included in the fits, with the predictions of our line lists and HITRAN. Table \ref{S} shows the comparison for absolute intensities. For the 0-0 band, all line lists work perfectly. The 1-0R(17) line is excellently reproduced by LL3 whereas LL1 and LL2 fail since they used the earlier 1-0 data excluded in LL3. The 3-0R(23) line as measured by Cygan \etal\ \cite{Cygan25} is reproduced by LL1 satisfactorily, LL2 and LL3 work even better; it should be stressed that all the line lists did not use these observational data, as well as the other ones mentioned in Tables \ref{S} and \ref{ratios}, in the fitting.

Table \ref{S} certifies that LL2 and LL3 show the overall significant improvement (excluding a few lines) with respect to HITRAN2020: 70\% of all lines in the LL3 column describe experiment within the experimental uncertainties. In particular, the 7-0 band is fully reproduced by LL3 with such an accuracy, which is due to the inclusion of these data in the fit. More significant are the results of LL2 for the 7-0 band calculated before the measurements. While not all of the lines are described within the experimental errors, the observed-minus-calculated differences are under 14\%, much less than the estimated theoretical uncertainty of 30\%. This means that the irregular DMF gives reasonable predictions for practical applications. Also noticeable is the significant improvement of LL3 against LL2 due to inclusion of the new high-precision data. The estimates of the calculated uncertainties in LL3 for higher overtones are given as well \cite{Ushakov25CO}, which can be of value for planning future experiments.

Balashov \etal\ \cite{Bailey25CO} set the task to construct an \ai\ DMF capable of \emph{describing} the measured intensities within the experimental uncertainties \emph{for all bands simultaneously}. They constructed a new \ai\ DMF and fitted it with a $6^\textrm{th}$-order polynomial, which solved in part this problem. Moreover, the necessity to \emph{predict} the yet unobserved lines in higher overtones was also formulated. Later, Zobov \etal\ \cite{Zobov25} created a more successful \ai\ DMF that gave ``the line intensities of all observed vibrational bands from 0–0 to 7–0 with the sub-percent or experimental accuracy (whichever is the larger)", as demonstrated in their Table 3\footnote{There are some misprints: the 0-0R(10,15,20) and 2-0P(10) line intensities (and reference to Bielska \etal\ (2022) as measured the 2-0 band, see Ref. \cite{Devi12erratum} instead) are not correct; the frequency and intensity of the 4-0P(22) are incorrect.}. Here, we have shown that our semi-empirical irregular DMF already solves both these tasks. 

The new method of Li \etal\ \cite{Li25ratios30} opened a possibility to greatly increase the accuracy by measuring the intensity ratios of pairs of lines. This permitted to reduce the experimental error by about 30 times, from 0.1 to 0.003\%. Table \ref{ratios} shows the comparison of the observations with LL3. Obviously, the observed-minus-calculated differences greatly exceed the experimental errors. The unprecedented low experimental error can be reached semi-empirically by refitting the DMF with a probable increase in the number of parameters. However, we don't see any necessity in such refitting since the differences between the calculated and observed intensity ratios are already extremely small, less than 0.1\% above the R(15)/P(15) ratio in the table and less than 0.2\% below it, which is more than sufficient for any practical tasks.

The \ai\ calculations of Li \etal\ \cite{Li25ratios30} reached the experimental accuracy for the intensity ratios, yet the analytic form of DMF was not presented. 
There are also a few \ai\ DMFs constructed by these and other authors \cite{Balashov23,Spirko24,Usov24,Meshkov24,Koput24,Koput25,Zobov25}, some in analytic forms, but the ability of the interpolated/extrapolated \ai\ DMFs to predict the intensities of yet unobserved lines has not been demonstrated, to the best of our knowledge. In contrast, our semi-empirical irregular and rational DMFs \cite{Ushakov25CO} provide for an acceptable level of accuracy for all the observed bands as well as for qualitative estimates of uncertainties of the calculations for higher overtones. This conclusion will be further verified when the new data \cite{Li25PRL} announced in the recent update of HITRAN \cite{Gordon25} are published.

\begin{figure}[htbp]
    \centering
    \includegraphics[width=0.75\linewidth]{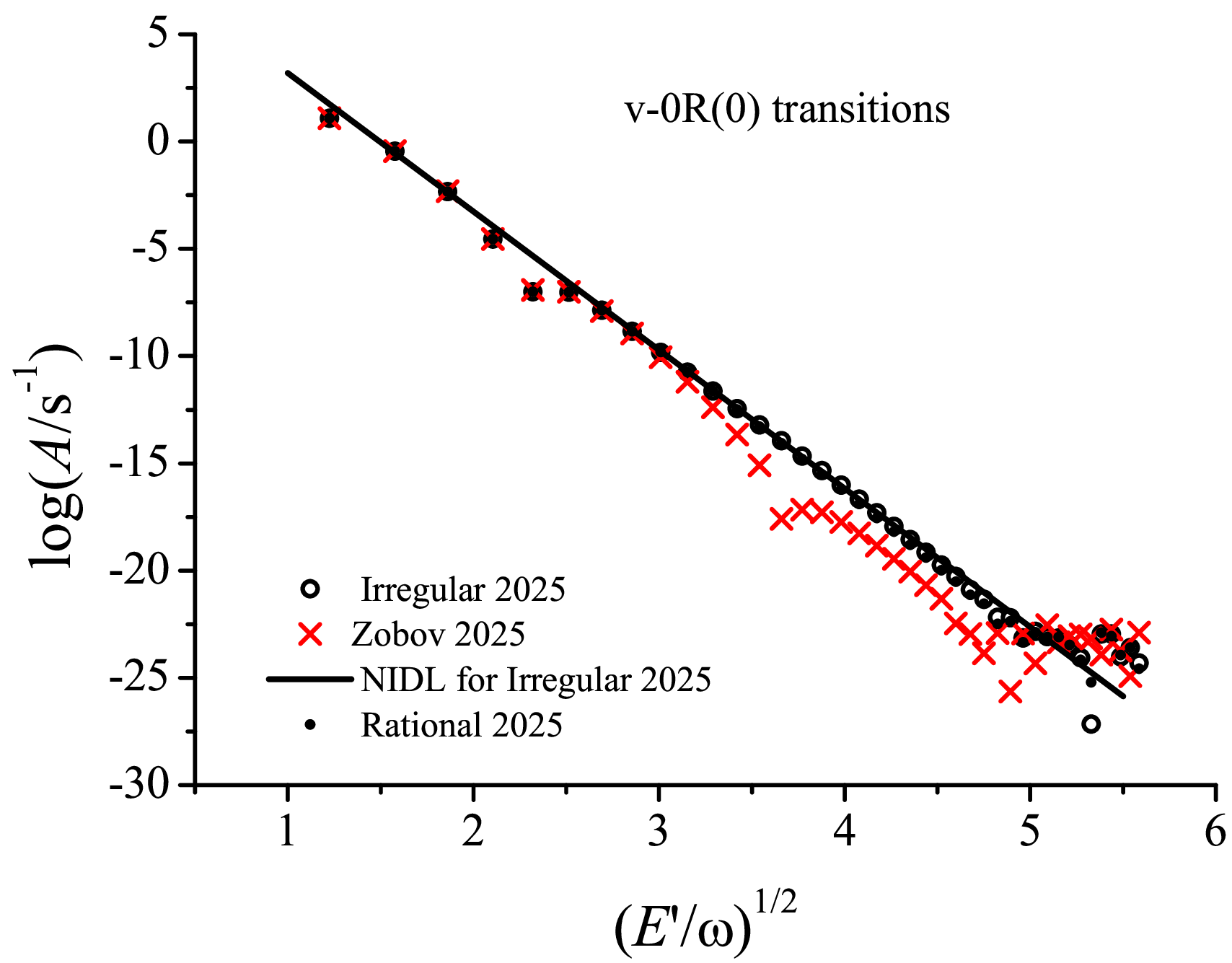}
    \caption{Comparison of the \EinA\ for the $v$-0R(0) lines calculated with the Irregular (empty circles) and Rational (full circles) DMFs of Ref. \cite{Medvedev22} fitted to the updated data base of Ref. \cite{Ushakov25CO} with the predictions of the polynomial DMF fitted to the \ai\ DMF in Ref. \cite{Zobov25} (crosses). $E^\prime$ is the energy of the upper level in cm$^{-1}$, $\omega=2143$ cm$^{-1}$. The NIDL line is drawn for the $v\le20$ overtone transitions excluding the 5-0 anomaly and two neighboring lines.}
    \label{fig}
\end{figure}

Figure \ref{fig} demonstrates the quality of predictions of our DMFs in comparison with the 6$^\textrm{th}$-order polynomial DMF of Zobov \etal\ (2025) \cite{Zobov25}. The \EinA\ for the $v$-0R(0) lines are plotted in the NIDL \cite{Medvedev12} coordinates. The NIDL line is drawn over $v=2,3,$7-20 excluding the anomaly at $v=5$ and two neighboring lines as well as the highest transitions requiring higher numerical precision. The NIDL line is perfectly straight, the absolute standard deviation for both DMFs being only 0.05 over all the transitions covered by the NIDL. In contrast, the intensities calculated by us with the polynomial of Zobov \etal\ do not obey the NIDL. Other above-cited analytic (fitted to \ai) DMFs demonstrate similar behavior, with the intensities variable among these DMFs by orders of magnitude; therefore, their predictions are not reliable. The latter conclusion is valid not only for the high-overtone transitions, which are mostly not observable, but also for the low $\Delta v$ transitions in view of the claimed percent \cite{Balashov23}, sub-percent \cite{Zobov25}, and even sub-promille \cite{Bielska22,Li25ratios30} accuracies of the calculations.

\begin{table}[htbp]
    \centering
    \caption{Comparison of the measured and predicted$^a$ absolute line intensities not included in the fitting of the semi-empirical DMFs}
    \vspace{7pt}
    \begin{tabular}{|c|c|c|c|c|c|c|c|}
    \hline
      &  &  & \multicolumn{4}{c|}{$\Delta=\textrm{100(O-C)/O}^\dagger $} & \\
      \hline
    line & $S_\textrm{obs}^b$ & Exp.uncert,\% & LL1 & LL2 & LL3$^c$  & HITRAN & source \\
    \hline
0-0R(0)   &   3.2814E-24   &   0.2   &   \textbf{0.006}   &   \textbf{0.006}   &   \textbf{0.005}   &   -0.6 & \cite{Tretyakov23} \\
0-0R(1)   &   2.553E-23   &   0.5   &   \textbf{0.09}   &   \textbf{0.09}   &   \textbf{0.08}   &   \textbf{-0.5} & \cite{Tretyakov23} \\
1-0R(17)   &   1.028E-19   &   0.6   &   1.5   &   1.5   &   \textbf{-0.4}   &   2.1 & \cite{Bailey25CO}$^d$ \\
3-0R(23)   &   8.0603E-25   &   0.09   &   \textbf{0.09}   &   \textbf{0.04}   &   \textbf{0.05}   &   0.17 & \cite{Cygan25}\\

3-0R(23)	&	8.1592e-025$^\star$	&	0.07	&	\textbf{-0.04}	&	-0.1	&	-0.08	&	\textbf{0.03}	&	\cite{Huang24},CRDS$^d$ \\
3-0R(25)	&	3.6455e-025	&	0.09	&	\textbf{-0.02}	&	\textbf{-0.07}	&	\textbf{-0.07}	&	\textbf{0.04}	&	\cite{Huang24},CRDS$^d$ \\
3-0R(26)	&	2.3679e-025	&	0.08	&	0.12	&	\textbf{0.07}	&	\textbf{0.06}	&	0.17	&	\cite{Huang24},CRDS$^d$ \\
3-0R(27)	&	1.5055e-025	&	0.07	&	0.1	&	\textbf{0.05}	&	\textbf{0.04}	&	0.08	&	\cite{Huang24},CRDS$^d$ \\
3-0R(28)	&	9.3917e-026	&	0.07	&	0.15	&	0.09	&	0.08	&	0.14	&	\cite{Huang24},CRDS$^d$ \\
3-0R(29)	&	5.7443e-026	&	0.07	&	0.17	&	0.12	&	0.1	&	0.16	&	\cite{Huang24},CRDS$^d$ \\
3-0R(30)	&	3.4436e-026	&	0.07	&	0.14	&	0.09	&	\textbf{0.07}	&	0.1	&	\cite{Huang24},CRDS$^d$ \\
3-0R(31)	&	2.0261e-026	&	0.07	&	0.18	&	0.12	&	0.1	&	0.09	&	\cite{Huang24},CRDS$^d$ \\
3-0R(32)	&	1.1684e-026	&	0.07	&	0.13	&	0.08	&	\textbf{0.05}	&	\textbf{0.06}	&	\cite{Huang24},CRDS$^d$ \\
3-0R(23)	&	8.1673e-025	&	0.08	&	\textbf{0.06}	&	\textbf{0.004}	&	\textbf{0.02}	&	0.13	&	\cite{Huang24},CMDS$^d$ \\
3-0R(25)	&	3.6501e-025	&	0.07	&	0.11	&	\textbf{0.05}	&	\textbf{0.06}	&	0.17	&	\cite{Huang24},CMDS$^d$ \\
3-0R(26)	&	2.3703e-025	&	0.07	&	0.22	&	0.17	&	0.17	&	0.27	&	\cite{Huang24},CMDS$^d$ \\
3-0R(27)	&	1.5069e-025	&	0.08	&	0.19	&	0.14	&	0.13	&	0.18	&	\cite{Huang24},CMDS$^d$ \\
3-0R(28)	&	9.3946e-026	&	0.07	&	0.18	&	0.12	&	0.11	&	0.17	&	\cite{Huang24},CMDS$^d$ \\
3-0R(29)	&	5.7481e-026	&	0.07	&	0.24	&	0.19	&	0.17	&	0.22	&	\cite{Huang24},CMDS$^d$ \\
3-0R(30)	&	3.4449e-026	&	0.08	&	0.18	&	0.13	&	0.11	&	0.13	&	\cite{Huang24},CMDS$^d$ \\
3-0R(31)	&	2.0268e-026	&	0.07	&	0.21	&	0.16	&	0.13	&	0.13	&	\cite{Huang24},CMDS$^d$ \\
3-0R(32)	&	1.1688e-026	&	0.09	&	0.16	&	0.11	&	\textbf{0.08}	&	\textbf{0.09}	&	\cite{Huang24},CMDS$^d$ \\

7-0P(19)   &   1.807E-30   &   21   &      &   \textbf{15}   &   \textbf{8.7}   &   44 & \cite{Balashov23} \\
7-0P(16)   &   3.645E-30   &   7.4   &      &   \textbf{6}   &   \textbf{-0.8}   &   42 & \cite{Balashov23} \\
7-0P(13)   &   6.698E-30   &   4.1   &      &   7   &   \textbf{0.3}   &   47 & \cite{Balashov23} \\
7-0P(12)   &   7.530E-30   &   4.0   &      &   \textbf{4}   &   \textbf{-3.1}   &   46 & \cite{Balashov23} \\
7-0P(11)   &   9.067E-30   &   4.1   &      &   10   &   \textbf{3.0}   &   50 & \cite{Balashov23} \\
7-0P(10)   &   9.592E-30   &   3.7   &      &   6   &   \textbf{-1.2}   &   49 & \cite{Balashov23} \\
7-0P(9)   &   1.065E-29   &   3.3   &      &   9   &   \textbf{2.4}   &   52 & \cite{Balashov23} \\
7-0P(8)   &   1.099E-29   &   3.4   &      &   9   &   \textbf{1.7}   &   52 & \cite{Balashov23} \\
7-0P(7)   &   1.103E-29   &   3.4   &      &   9   &   \textbf{1.7}   &   53 & \cite{Balashov23} \\
7-0P(4)   &   8.445E-30   &   3.8   &      &   8   &   \textbf{1.2}   &   55 & \cite{Balashov23} \\
7-0P(1)   &   2.543E-30   &   11   &      &   14   &   \textbf{6.6}   &   58 & \cite{Balashov23} \\
7-0R(0)   &   2.462E-30   &   12   &      &   \textbf{12}   &   \textbf{4.4}   &   58 & \cite{Balashov23} \\
7-0R(4)   &   9.119E-30   &   3.5   &      &   7   &   \textbf{-1.3}   &   56 & \cite{Balashov23} \\
7-0R(6)   &   9.988E-30   &   3.5   &      &   6   &   \textbf{-2.4}   &   55 & \cite{Balashov23} \\
\hline
\multicolumn{8}{l}{$^\dagger$The $\Delta$'s falling within the experimental uncertainties are shown in bold face.}\\
\multicolumn{8}{l}{$^\star$ 8.0494 for natural abundance, differs from Ref. \cite{Cygan25} by 0.13\%.}\\
\multicolumn{8}{l}{$^a$Recalculated to HITRAN's natural abundance where necessary.}\\
\multicolumn{8}{l}{$^b$For 296 K and HITRAN's natural abundance if not indicated otherwise.}\\
\multicolumn{8}{l}{$^c$The 7-0 data were used in the fitting.}\\
\multicolumn{8}{l}{$^d$100\% abundance. CRDS, \emph{absorption}; CMDS, \emph{dispersion}.} 
    \end{tabular}
    \label{S}
\end{table}

\begin{table}[htbp]
    \centering
    \caption{Comparison of measured by Li \etal\ \cite{Li25ratios30} line 1 to line 2 intensity ratios, S1/S2, at various temperatures and predicted by LL3 relative line intensities in the 3-0 band recalculated to the experimental temperature}
    \vspace{7pt}
    \begin{tabular}{|c|c|c|c|c|c}
    \hline
L1/L2   &   T,K   &   S1/S2   & Exp. uncert,\%  &  100(O-C)/O   \\

\hline
R0/R1   &   298.429   &   0.503012   &   0.004   &   -0.007   \\
R1/R3   &   298.430   &   0.534955   &   0.003   &   -0.007   \\
R3/R5   &   298.426   &   0.767994   &   0.002   &   -0.013   \\
R5/R10   &   298.431   &   1.073886   &   0.002   &   -0.013   \\
R1/R10   &   298.431   &   0.441206   &   0.002   &   -0.031   \\
R10/R13   &   298.430   &   1.472357   &   0.001   &   -0.012   \\
R13/R15   &   298.429   &   1.458483   &   0.004   &   -0.006   \\
R15/R17   &   298.426   &   1.594931   &   0.003   &   -0.006   \\
R17/R19   &   298.430   &   1.738204   &   0.002   &   -0.005   \\
R19/R21   &   298.430   &   1.889597   &   0.002   &   -0.009   \\
R21/R22   &   298.432   &   1.417658   &   0.002   &   0.005   \\
R22/R23   &   298.434   &   1.446427   &   0.004   &   -0.008   \\
R23/R24   &   298.437   &   1.475842   &   0.0014   &   0.0007   \\
R24/R25   &   298.434   &   1.505442   &   0.003   &   -0.004   \\
R25/R26   &   298.429   &   1.535529   &   0.003   &   -0.002   \\
R26/R27   &   298.428   &   1.566004   &   0.0040   &   -0.0008   \\
R1/P1   &   298.428   &   2.076074   &   0.006   &   0.008   \\
R5/P5   &   298.430   &   1.376124   &   0.002   &   0.040   \\
R5/P5   &   298.429   &   1.376132   &   0.003   &   0.040   \\
R5/P5   &   298.928   &   1.376095   &   0.003   &   0.038   \\
R10/P10   &   298.428   &   1.428264   &   0.003   &   0.068   \\
R15/P15   &   298.428   &   1.567518   &   0.003   &   0.099   \\
\hline
R17/P17   &   298.427   &   1.634831   &   0.004   &   0.117   \\
R20/P20   &   298.428   &   1.745511   &   0.004   &   0.140   \\
R21/P21   &   298.427   &   1.784819   &   0.005   &   0.145   \\
R21/P21   &   298.430   &   1.784829   &   0.002   &   0.146   \\
R21/P21   &   298.428   &   1.784802   &   0.003   &   0.144   \\
R23/P23   &   298.431   &   1.867087   &   0.003   &   0.161   \\
R23/P23   &   298.427   &   1.867099   &   0.004   &   0.161   \\
R23/P23   &   298.427   &   1.867157   &   0.003   &   0.164   \\
R23/P23   &   298.429   &   1.867182   &   0.003   &   0.166   \\
R26/P26   &   298.427   &   1.999426   &   0.007   &   0.189   \\
R26/P26   &   298.428   &   1.999467   &   0.004   &   0.191   \\
R26/P26   &   298.431   &   1.999478   &   0.005   &   0.191   \\
R26/P26   &   298.428   &   1.999611   &   0.003   &   0.198   \\
R26/P26   &   298.429   &   1.999612   &   0.003   &   0.198   \\
R26/P26   &   298.428   &   1.999598   &   0.003   &   0.197   \\
R26/P26   &   298.427   &   1.999587   &   0.003   &   0.197   \\
    \hline
    \end{tabular}
    
    \label{ratios}
\end{table}

\section*{Acknowledgement}

This work was performed under state task, Ministry of Science and Education, Russian Federation, state registration number 124013000760-0.

\newpage
\bibliography{ESMedvedev-Recovered-2,Overtones-Converted-Recovered-2}
\bibliographystyle{elsarticle-num}
\end{document}